\documentclass[conference]{IEEEtran}
\ifCLASSOPTIONcompsoc
  \usepackage[nocompress]{cite}
\else
  \usepackage{cite}
\fi
\usepackage{graphicx}
\usepackage{url}
\usepackage{amsmath}
\ifCLASSINFOpdf
\else
\fi
\usepackage{booktabs}
\begin{document}
%
\title{Transformer-Gather, Fuzzy-Reconsider: \\A Scalable Hybrid Framework for Entity Resolution}


\author{
\IEEEauthorblockN{Mohammadreza Sharifi}
\IEEEauthorblockA{Department of Computer Engineering\\
Ferdowsi University of Mashhad,\\ Mashhad 9177948974, Iran\\
sharifi.mohammadreza@mail.um.ac.ir}
\IEEEauthorblockA{Iranserver Inc.
Mashhad 9185837715, Iran\\
mr.sharifi@iranserver.com}

\and

\IEEEauthorblockN{Danial Ahmadzadeh}
\IEEEauthorblockA{Iranserver Inc.\\
Mashhad 9185837715, Iran\\
d.ahmadzadeh@iranserver.com}
}


%


\IEEEoverridecommandlockouts
\IEEEpubid{%
  \makebox[\columnwidth]{979-8-3315-8973-8/25/\$31.00~\copyright~2025 IEEE\hfill}
  \hspace{\columnsep}\makebox[\columnwidth]{}%
}
\maketitle

\IEEEpubidadjcol

\begin{abstract}
Entity resolution plays a significant role in enterprise systems where data integrity must be rigorously maintained. Traditional methods often struggle with handling noisy data or semantic understanding, while modern methods suffer from computational costs or the excessive need for parallel computation. In this study, we introduce a scalable hybrid framework, which is designed to address several important problems, including scalability, noise robustness, and reliable results. We utilized a pre-trained language model to encode each structured data into corresponding semantic embedding vectors. Subsequently, after retrieving a semantically relevant subset of candidates, we apply a syntactic verification stage using fuzzy string matching techniques to refine classification on the unlabeled data.  
This approach was applied to a real-world entity resolution task, which exposed a linkage between a central user management database and numerous shared hosting server records. Compared to other methods, this approach exhibits an outstanding performance in terms of both processing time and robustness, making that a reliable solution for a server-side product. Crucially, this efficiency does not compromise results, as the system maintains a high retrieval recall of approximately 0.97. The scalability of the 'Transformer-Gather, Fuzzy-Reconsider' framework makes it deployable on standard CPU-based infrastructure, offering a practical and effective solution for enterprise-level data integrity auditing.
\end{abstract}
%
\IEEEpeerreviewmaketitle

\section{Introduction}
In a modern enterprise system, maintaining consistency across disparate sources of data is essential \cite{loshin2010master}. For service providers, such as web hosting companies, user data is frequently distributed over a billing system, management database, and numerous operational service endpoints, like shared hosting servers. Over time, records in these systems inevitably diverge due to specific reasons, including manual data modification, network connection Problems, and differing update cycles. A single user may face a variety of contradictions, including different email, username, status of billing, website domain, and the hosting cluster name. These types of contradictions provide a massive amount of friction for some teams, including operational support and customer care. That is the main reason we conducted to find a robust solution to prevent unwanted changes and termination for the users.

The entity-resolution (ER) is the task of identifying records that correspond to the same real-world entity across diverse data sources.
This key challenge in data management poses a significant obstacle, which described earlier. Old conventional methods typically employ deterministic, rule-driven matching confined to narrow or fixed search spaces. Despite the effectiveness of these approaches, they are extremely vulnerable to the scenario of noisy data, which may result in some undemanding changes to the user's information. Furthermore, using these types of ideas could be significantly expensive to play a role in a CPU server's product.
As we face an approximation comparison on a couple of large datasets, we need to endure the complexity of $O(nm)$, which is huge for a non-parallelizable product.

Recent advances in deep learning and especially natural language processing have shown some beneficial impacts on ER problems \cite{barlaug2021neural}. By leveraging transformer-based models, we will explore a new avenue in the field of semantic understanding of entities. However, relying solely on the semantic phase potentially leads us to a devastating narrow cliff of reliability. Especially when struggling with real-world customers, we are in search of reliable answers that should include both semantic and syntactic phases. 

To address these limitations, we introduce a framework, Transformer-Gather, Fuzzy-Reconsider (TGFR), to provide a scalable and hybrid approach for the entity resolution problem. Our approach consisted of a two-stage pipeline that can mix the strength of transformers alongside the precision of the traditional methods. As we need reliable results, we use the deep semantic understanding of a pre-trained sentence transformer to encode the well-structured entities. Alongside the semantic understanding, we utilize a retrieval method to narrow down the searching space, which is subsequently passed to an approximate matching layer to tag the unlabeled data properly. 

This framework has been deployed in a production environment for over a year, serving as the core engine for a large-scale data integrity project. The system's effectiveness is demonstrated by maintaining a high retrieval recall of approximately 0.97 while achieving low-latency processing times on standard CPU-based infrastructure, as most production environments don't meet the on-demand GPU needs, it's a key factor for a framework to rely mostly on CPU. This performance validates the framework as an applicable and reliable solution for classifying relationships within large, unlabeled databases, as demonstrated in its application at the Iranserver company.

\section{Related Works}
Early ER methods mainly discussed deterministic solutions and exact matching approaches \cite{hernandez1995}, which are compromised by noisy data. Furthermore, approximate string matching ideas\cite{navarro2001guided} (e.g., Levenshtein \cite{levenshtein1966binary}) were able to improve the results in terms of ER problems \cite{cohen2003comparison,ristad2002learning} with a deficiency of semantic understanding. Meanwhile, by leveraging classic machine learning approaches \cite{bilenko2003adaptive, yujian2007normalized}, ER problems faced with substantial results \cite{ bahmani2017erblox, wang2011entity} that heavily relied on handcrafted features. By the enormous impact of deep learning models \cite{ebraheem2017deeper} and especially transformers, ER problems are recognized as yielding beneficial outcomes. By using transformer-based methods\cite{li2021deep} and specially pre-trained language models (e.g. BERT \cite{BERT} ) in ER problems, robust and significant results were encountered \cite{langmodel_ditto}. However, their computational cost limits scalability in the action. Hybrid systems with two-stage, retrieve and re-rank architecture \cite{vassilis}, balance scalability and accuracy while maintaining the robustness to noisy data \cite{sun2021entity}. Recently, Zero-shot classifiers have shown promising results \cite{wu2020zeroer}, but ultimately lack detection speed, which is important for server-side products that mostly are not able to access an on-demand GPU. Our Transformer-gather, Fuzzy-reconsider (TGFR) framework not only enhances scalability and robustness but, provides a significantly rapid and deployable approach for implementation on servers that do not require GPU access.

\section{Methodology}
In this study, we have conducted a pipeline that subsequently encodes, retrieves, and finally matches the entities across relevant datasets. The methodological components of this hybrid framework are detailed subsequently.

\subsection{Dataset}
The datasets for this study were gathered from diverse sources of data belonging to Iranserver, a subsidiary of the Greenweb company, which operates in the fields of web hosting and cloud infrastructure services.
The main database cluster provides a sufficient number of related fields, including Email, Username, Domain, Billing Status, and Cluster name. Thus, we can use these extracted data as the reference set that should be matched with a variety of users distributed across a massive number of clusters. 
The second dataset, as mentioned earlier, involves the pivotal data of the users that were used to clarify the contradictions. These data include the same fields as the reference dataset, as described above. By collecting some substantial data, which shapes a couple of huge datasets, we are now prepared to advance further using transformers.

\subsection{Data Preparation and Encoding}
In order to encode each entry of the datasets, we need a simple format to transform each row into a sentence. This sentence could be shaped as a simple linguistic prompt that includes important information fields.
\begin{figure}[h!]
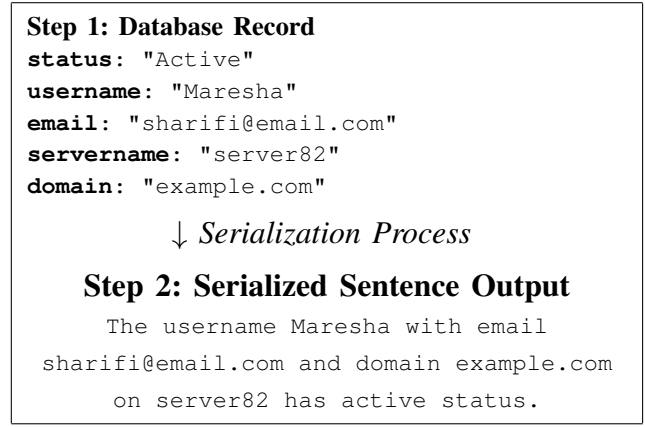

\centering 
\begin{tabular}{|l|} 
\hline
\begin{minipage}{0.9\columnwidth} 
\vspace{0.5em} 

\textbf{Step 1: Database Record} \\
\texttt{\small \textbf{status}: "Active"} \\
\texttt{\small \textbf{username}: "Maresha"} \\
\texttt{\small \textbf{email}: "sharifi@email.com"} \\
\texttt{\small \textbf{servername}: "server82"} \\
\texttt{\small \textbf{domain}: "example.com"} 

\vspace{0.5em} 
\centering \large ${\downarrow}$ \textit{Serialization Process} 
\vspace{0.5em} 
\\

\textbf{Step 2: Serialized Sentence Output} \\
\texttt{\small The username Maresha with email sharifi@email.com and domain example.com on server82 has active status.} \\
\vspace{0.5em} 
\end{minipage} \\
\hline
\end{tabular}
\caption{An example of serializing a structured user record into a standardized sentence format.}
\label{fig:sentence_creation} 
\end{figure}

As each row of the dataset is properly serialized, we leverage a \textit{sentence-transformer} model to significantly enhance the matching and classification performance \cite{reimers2019sentence}. In this study, we utilize the pre-trained \textit{distilbert-base-uncased} model, which retains over 95\% of BERT’s performance while offering substantial improvements in inference speed and memory usage \cite{sanh2019distilbert}. After encoding the textual inputs into fixed-size embedding vectors using this bi-encoder architecture, we apply fast retrieval followed by re-ranking and approximate string matching techniques to identify highly accurate matches across records.
\label{datapreparation}

\subsection{Retrieval Phase}
By leveraging an appropriate embedding that was provided by the model, we expect similar sentences with similar information to be close to each other. Since the current search space per entity is much higher than the computational resources, we are not able to apply the exhaustive search in the $O(nm)$ time complexity. Meanwhile, this framework should provide high efficiency as there are no on-demand powerful parallelization resources (GPU) on the product server.
Thus, we are not able to use techniques like cross-encoding \cite{reimers2019sentence} for the initial candidate retrieval phase.

Therefore, a method to efficiently reduce the search space is required. To this end, we employ a K-Nearest Neighbors (KNN) retrieval technique \cite{cunningham2021k}. KNN allows for an efficient index-based search over the vector space, enabling the rapid identification of a small subset of the most semantically similar candidates for each query record. This part transforms the intractable problem of comparing millions of pairs into a manageable task of analyzing only a small, highly relevant candidate set.

\subsection{Approximate Matching}
\begin{figure}[!t]
    \centering
    \includegraphics[width=0.99\columnwidth]{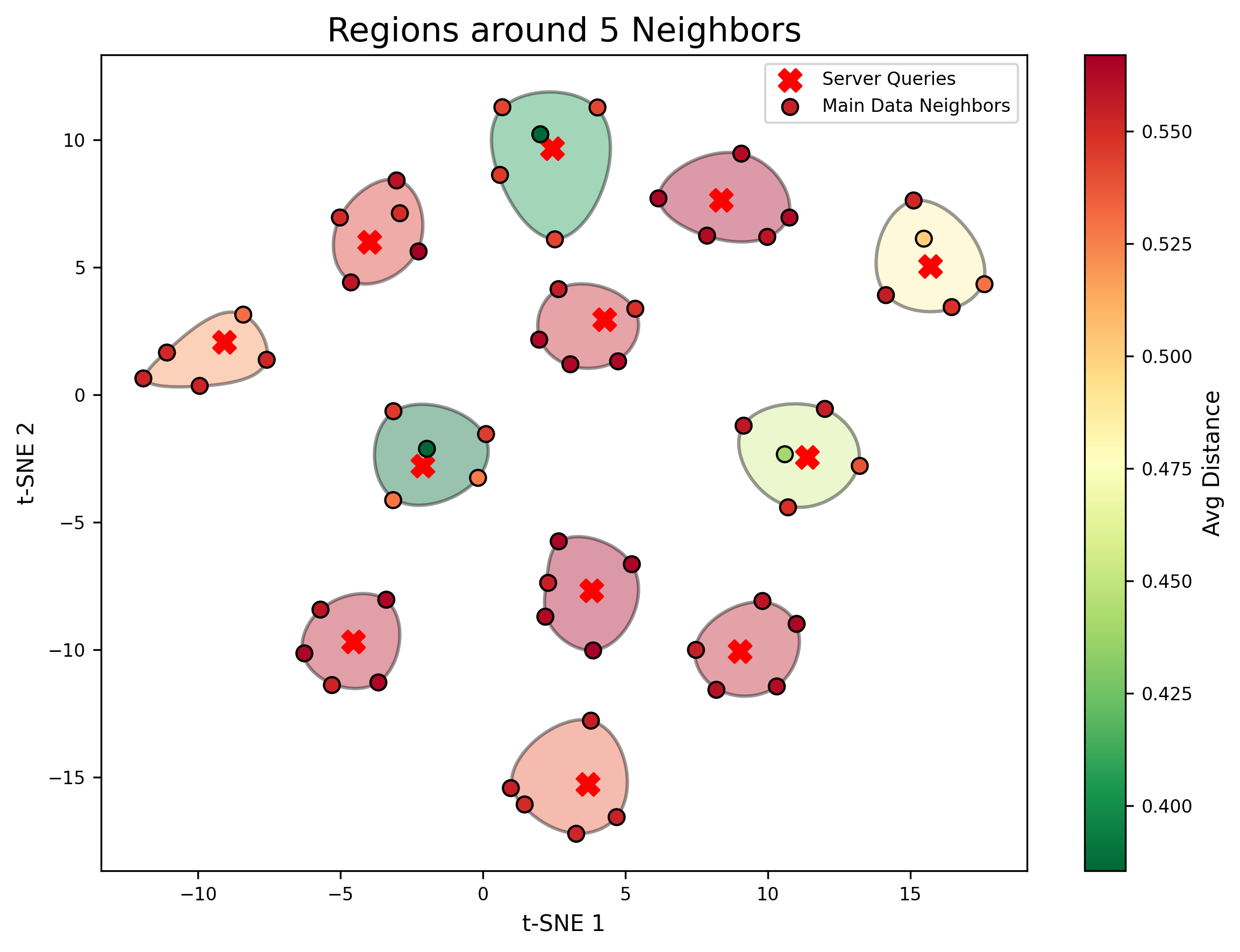}
    \caption{Example t-SNE illustration of matching regions formed by 12 sample server query and their $5$ nearest neighbors. The color intensity reflects the average cosine distance of the territory members to the main server data.}
    \label{fig:regions}
\end{figure}

By fitting a KNN model on the main database data, top 5 nearest neighbor to each query are now provided. Since the neighbors that forms a region might be very close to each other, as shown in Fig.~\ref{fig:regions}, we need to exclusively match them in the string domain rather than vector domain.

Thus, in order to perform an accurate matching, we leverage an approximate string matching, Levenshtein distance, to rate the similarity between the candidates and the query data.
To determine the best match from a retrieved candidate set, a composite score is calculated for each pair $(R_A, R_B)$, where $R_A$ is a source record and $R_B$ is a candidate. The score is a weighted average of the fuzzy string similarities over a set of key fields, $F = \{f_1, f_2, \dots, f_k\}$.

The final composite score, $\text{Score}(R_A, R_B)$, is defined as:

\begin{equation}
\label{eq:composite_score}
\text{Score}(R_A, R_B) = \sum_{i=1}^{k} w_i \cdot S(R_A(f_i), R_B(f_i))
\end{equation}

\noindent where:
\begin{itemize}
    \item $S(s_1, s_2)$ is a normalized Levenshtein distance function with a range of $[0, 1]$.
    \item $R_A(f_i)$ is the value of field $f_i$ in record $R_A$.
    \item $w_i$ is the empirically assigned weight for field $f_i$, reflecting its importance, such that $\sum_{i=1}^{k} w_i = 1$.
    \item For our specific implementation, the weights were assigned empirically to prioritize core identifiers: 
$w_{\text{email}}=0.4$, $w_{\text{username}}=0.3$, $w_{\text{domain}}=0.15$, $w_{\text{servername}}=0.1$, and $w_{\text{status}}=0.05$.

\end{itemize}

The candidate pair with the highest final score is selected as the best match for the subsequent tagging process. 

\subsection{Providing a Ground-Truth}
As we struggle with an unlabeled dataset, we need first to provide a ground-truth to have a meaningful comparison. Thus, we perform an exhaustive $O(nm)$ search using GPU alongside another approximate matching technique to find the linkable pairs and tag them properly.
We encoded all the possible entities using a pre-trained version of \textit{all-mpnet-base-v2} \cite{reimers2019sentence}, which was subsequently fine-tuned on our dataset.  To fine-tune our encoder, we employed the MNR-Loss \cite{reimers2019sentence}, a contrastive loss function that effectively learns sentence embeddings for semantic similarity. This loss treats each (anchor, positive) pair in a batch as the true match and all other sentences in the batch as implicit negatives.

Given a batch of $N$ pairs $(a_i, p_i)$ where $a_i$ is an anchor and $p_i$ is its corresponding positive, the loss for each anchor-positive pair is computed as:

\[
\mathcal{L}_{i} = -\log \frac{\exp(\text{sim}(a_i, p_i))}{\sum_{j=1}^{N} \exp(\text{sim}(a_i, p_j))}
\]

where $\text{sim}(\cdot, \cdot)$ is the cosine similarity between embeddings. The final loss is the average over all anchors in the batch:

\[
\mathcal{L} = \frac{1}{N} \sum_{i=1}^{N} \mathcal{L}_i
\]

\begin{figure}[h]
    \centering
    \includegraphics[width=0.9\linewidth]{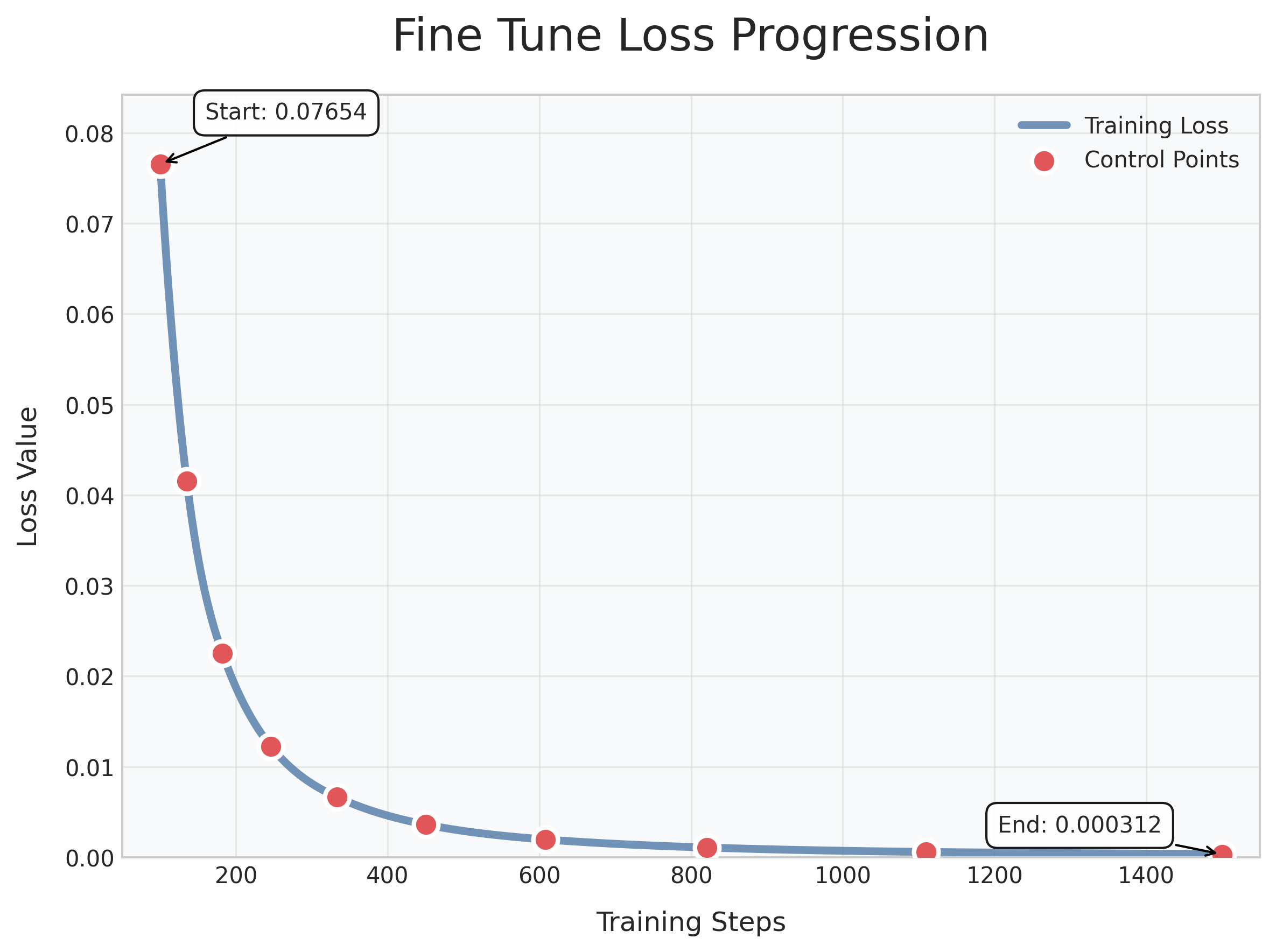}
    \caption{Training loss over steps curve. This curve shows a smooth training process and a low final loss value.}
    \label{fig:loss-curve}
\end{figure}

This formulation allows for efficient batch training by using all other positives as negatives, improving convergence and performance.
Figure~\ref{fig:loss-curve} illustrates the trend of training loss over time.
After the fine-tuning and encoding, we used the Cosine similarity to simultaneously gather all possible pairs having a threshold greater than $0.8$. This value is selected based on the business logic behind the datasets. Further, we ran a tagging mechanism to tag all the possible pairs appropriately.

\section{Experimental Results}
In this section, we present the results of this study in comparison to other known methods. All of these experiments were conducted on a server with an RTX3090 (24GB VRAM) GPU, Intel Xeon W-1245 CPU, and 96GB of RAM.

\subsection{Recall Performance of KNN}
\begin{figure}[h]
    \centering
    \includegraphics[width=0.99\linewidth]{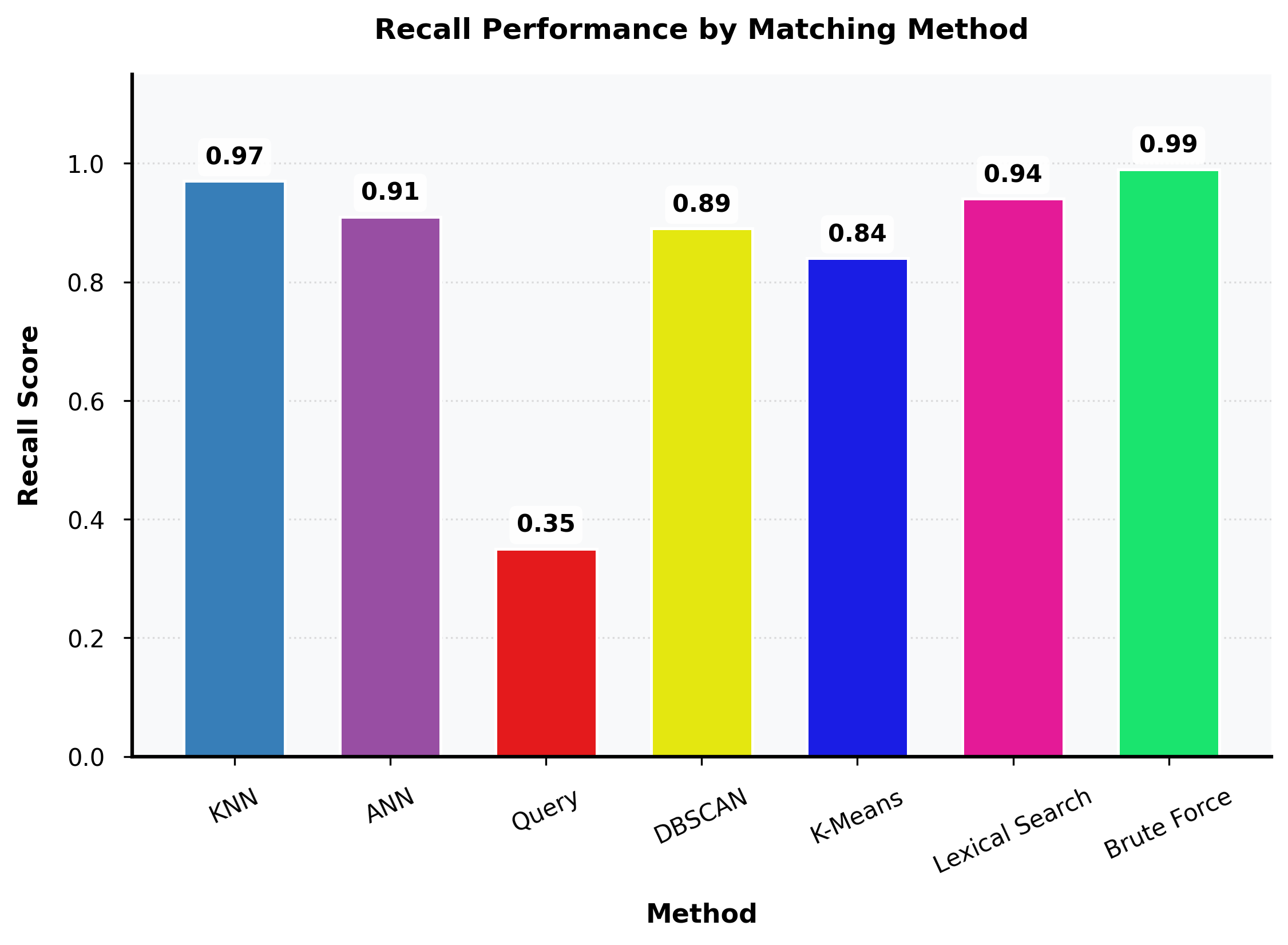}
    \caption{Recall values comparison for different methods. Left to right: KNN, ANN, Query matching, DBSCAN+ANN, K-means+ANN, Lexical Search, and Brute Force.}
    \label{fig:recalls_main}
\end{figure}

The recall metric is an important measurement that gauges the contribution of the framework in finding the contradictions. Since a false classification could result in a devastating outcome for the customers, we need to keep the recall as high as possible. Thus, we need a highly reliable way to search and narrow down the search space. Figure \ref{fig:recalls_main} illustrates a comparison plot of their recall metrics.

As we see, the KNN approach achieves a considerable performance on our data. Other methods, including ANN\cite{johnson2019billion}, may achieve faster retrieval speed. But, the problem is we need to determine a balance between the reliability of results and CPU-side retrieval latency. The execution time ratio compared to the Brute-Force is shown in the Table \ref{tab:speedup}.  
\begin{table}[htbp]
\caption{Execution Time Ratio Across Retrieval Methods}
\centering
\begin{tabular}{lr}
\toprule
\textbf{Method} & \textbf{Execution Time Ratio} \\
\midrule
Joint Query         & 13.83 \\
K-Means ANN                  & 7.23  \\
Lexical Search        & 4.13  \\
KNN (Vector Search)          & 4.10  \\
DBSCAN ANN                   & 1.51  \\
Brute-Force & 1.00  \\
\bottomrule
\end{tabular}
\label{tab:speedup}
\end{table}
The experimental results demonstrate that KNN achieves an optimal balance between recall and computational efficiency. While Joint Query offers the highest speedup, its recall performance falls significantly. ANN methods, despite their scalability advantages \cite{johnson2019billion}, exhibit a considerable recall drop compared to exact KNN, which is essential if missed contradictions carry high costs. The KNN approach provides a substantial speedup over Brute-Force while maintaining 97.2\% recall. This performance stems from KNN's exhaustive search in reduced candidate spaces and deterministic results, unlike ANN's approximations.

\subsection{Effect of Using a Fine-Tuned Model}
\begin{figure}[h]
    \centering
    \includegraphics[width=0.9\linewidth]{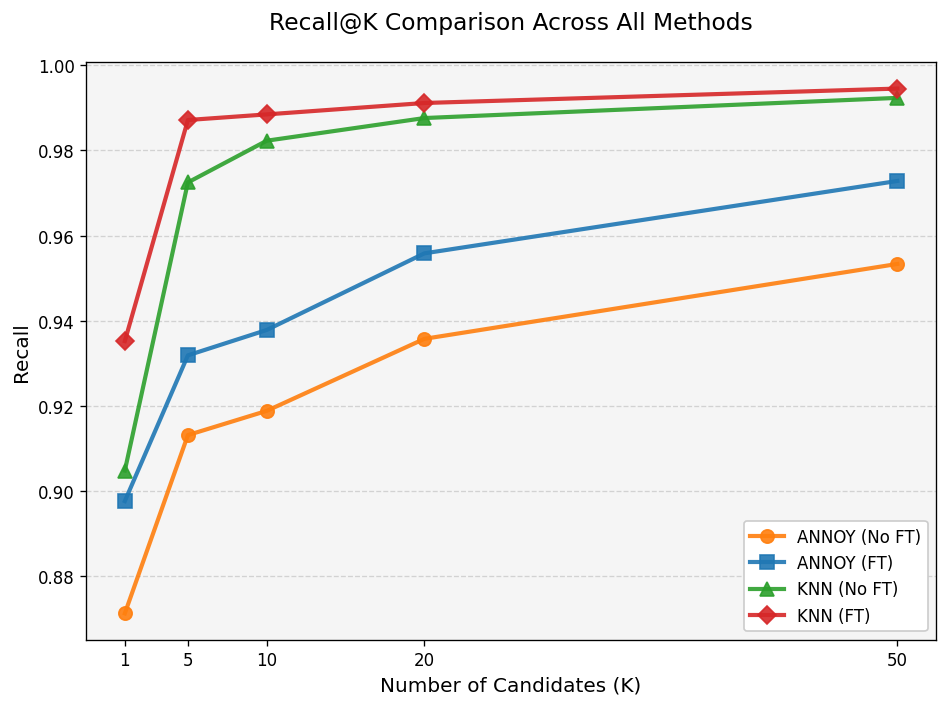}
    \caption{Recall comparison between KNN and ANNOY in both setups with fine-tuning or without fine-tuning.}
    \label{fig:pretrained}
\end{figure}

As mentioned earlier(Section \ref{datapreparation}), we used an off-the-shelf pretrained model, without any task-specific fine-tuning. By leveraging a simple fine-tuning step, we can achieve better results as the model can encode sentences into more distinguishable vectors.
In this experiment, we compared the effect of using a fine-tuning step on the recall results of ANNOY \cite{schutze2008introduction} and KNN with some specific K-values from 1 to 50. The Fig. \ref{fig:pretrained} illustrates the recall performance as described.

The impact of using a simple one fine-tuning step is extremely positive. Thus, we can apply the fine-tuning once and save the fine-tuned version of the model and use that appropriately. The detailed comparison is shown in Table \ref{tab:recall_at_k}.

\begin{table}[htbp]
\caption{Recall@K Comparison Across Retrieval Methods}
\centering
\begin{tabular}{lccccc}
\toprule
\textbf{Method} & \textbf{K=1} & \textbf{K=5} & \textbf{K=10} & \textbf{K=20} & \textbf{K=50} \\
\midrule
ANNOY (No FT)         & 0.8712 & 0.9131 & 0.9188 & 0.9357 & 0.9533 \\
ANNOY (FT)            & 0.8976 & 0.9319 & 0.9378 & 0.9558 & 0.9728 \\
KNN (No FT)           & 0.9048 & 0.9725 & 0.9823 & 0.9876 & 0.9923 \\
KNN (FT)              & 0.9352 & 0.9871 & 0.9885 & 0.9911 & 0.9945 \\
\bottomrule
\end{tabular}
\label{tab:recall_at_k}
\end{table}
The experimental results demonstrate significant improvements leveraging a single fine-tuning step. By using a single fine-tuning procedure, we can expect the model to create more semantically distinguishable vectors. These noticeable results bring the attractiveness of reaching an approximate recall score in comparison to brute-force. Since the KNN is faster than the brute-force matching technique, it sounds like a promising method. But we were faced with a faster approach by using ANNOY. The ANNOY framework is highly scalable and much faster than KNN, especially in the matter of big data \cite{johnson2019billion}. But, in our case of usage, we weren't struggling with a huge amount of data. Since the current KNN methodology has achieved a considerable performance, preferring a reliable framework could retain the benefits of having a balanced trade-off between result preparation latency and the metrical performances.

\subsection{Using a Transformer-based Model}
\begin{figure}[h]
    \centering
    \includegraphics[width=0.9\linewidth]{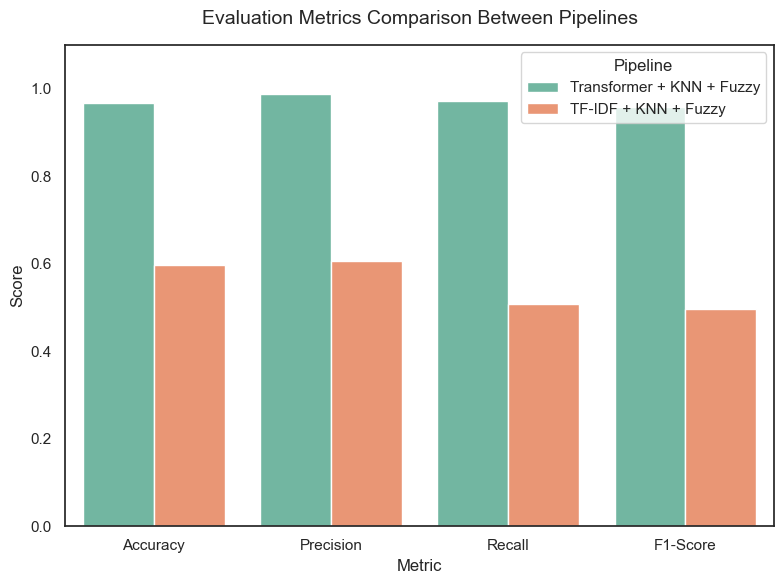}
    \caption{Impact of transformer-based encoding versus TF-IDF. }
    \label{fig:tfidf}
\end{figure}

Using a transformer-based encoder model could be significantly beneficial to have a reliable outcome. Our implementation leverages a pre-trained language model to generate contextual embeddings of user records, enabling semantic matching beyond syntactic similarity. Replacing this key part of the framework via older methods, like TF-IDF\cite{alpaydin2020introduction,schutze2008introduction}, could lead to unwanted results, as illustrated in Fig. \ref{fig:tfidf}.

Also, detailed values are shown in Table \ref{tab:pipeline_performance}.

\begin{table}[htbp]
\caption{Performance Comparison of Entity Resolution Pipelines}
\centering
\begin{tabular}{lcccc}
\toprule
\textbf{Pipeline} & \textbf{Accuracy} & \textbf{Precision} & \textbf{Recall} & \textbf{F1-Score} \\
\midrule
Transformer & 0.9682 & 0.9871 & 0.9725 & 0.9780 \\
TF-IDF     & 0.5977 & 0.6056 & 0.5090 & 0.4971 \\
    \bottomrule

\end{tabular}
\label{tab:pipeline_performance}
\end{table}
These results indicate that the effectiveness of using a transformer-based encoder is essential to having a reliable result. It significantly outperforms in all the mentioned metrics. These considerable results show the effectiveness of having a proper embedding vector, as they represent each entity with sufficient semantic density to preserve relationships while maintaining discriminative power. These considerable improvements demonstrate the limitations of TF-IDF's bag-of-words representations, which fail to capture semantic relationships and contextual nuances that transformer embeddings preserve.

\subsection{Impact of Fuzzy String Approximation}
\label{subsec:fuzzy_impact}

\begin{figure}[htbp]
    \centering
    \includegraphics[width=0.9\linewidth]{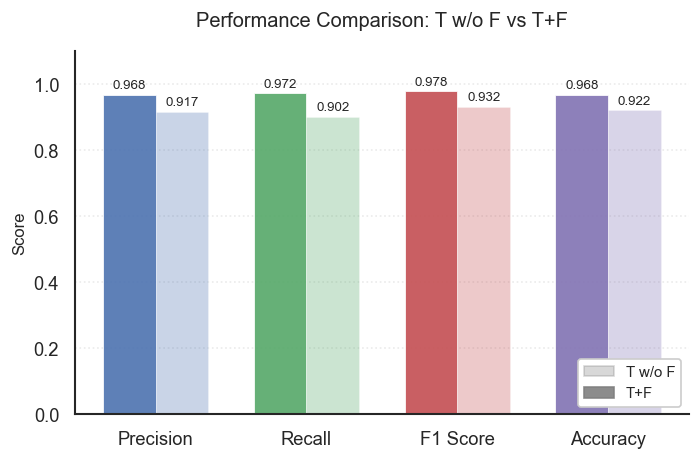}
    \caption{Performance comparison between transformer-only (T w/o F) and transformer with fuzzy matching (T+F) approaches.}
    \label{fig:t_comparison}
\end{figure}

The addition of fuzzy string matching as a final verification layer demonstrates significant improvements in entity resolution quality. Figure~\ref{fig:t_comparison} compares the performance of the transformer-only approach (T w/o F) versus the full pipeline with fuzzy matching (T+F), while Table~\ref{tab:transformer_comparison} provides detailed metric comparisons.

\begin{table}[htbp]
    \caption{Performance Metrics With and Without Fuzzy Matching}
    \centering
    \begin{tabular}{lcccc}
    
\toprule
    \textbf{Method} & \textbf{Precision} & \textbf{Recall} & \textbf{F1 Score} & \textbf{Accuracy} \\
    \midrule
    T w/o F & 0.9166 & 0.9017 & 0.9320 & 0.9217 \\
    T+F & 0.9682 & 0.9722 & 0.9780 & 0.9682 \\
    \bottomrule
    \end{tabular}
    \label{tab:transformer_comparison}
\end{table}
The fuzzy matching layer improves performance by correcting typographical mismatches that semantic models may miss, normalizing format variants using character-level similarity, and preserving high precision while enhancing recall, thus serving as a reliable verification layer rather than a primary matcher.
The experimental results demonstrate that fuzzy matching effectively complements transformer-based semantic matching, successfully bridging the gap between high-level semantic understanding and precise string-level matching. The observed 4.6 percentage point improvement in F1-score (from 0.932 to 0.978) confirms that our hybrid approach captures both semantic relationships and syntactic variations more comprehensively than either method could achieve independently. This dual capability proves particularly valuable in enterprise environments, where we must handle frequent data entry variations while maintaining rigorous precision standards - a balance that single-method approaches consistently fail to achieve. Since an unreliable result could lead to termination or unwanted automated actions, we need some critical steps like this to prevent invalid actions. 

\subsection{Scalability Analysis}

Our hybrid pipeline achieves efficient scaling through three optimized stages. Let $m$ be source records, $n$ target records, $d = 768$ (embedding dimension), $k = 5$ (top candidates), and $L = 25$ (verification length):

\begin{table}[ht]
\centering
\caption{Per-Stage Computational Complexity}
\begin{tabular}{lcc}
\toprule
\textbf{Stage} & \textbf{Our Approach} & \textbf{Brute-Force} \\
\midrule
Encoding & $O((n + m)d^2)$ & $O((n + m)d^2)$ \\
Search & $O(n\log m)$ & $O(nm)$ \\
Verification & $O(nkL)$ & $O(nmL)$ \\
\bottomrule
\end{tabular}
\label{tab:stage_complexity}
\end{table}

\begin{table}[ht]
\centering
\caption{Total Computational Complexity}
\begin{tabular}{lc}
\toprule
\textbf{Approach} & \textbf{Complexity} \\
\midrule
Our Hybrid Pipeline & $O((n + m)d^2 + n(\log m + kL))$ \\
Brute-force Baseline & $O((n + m)d^2 + nm(1 + L))$ \\
\bottomrule
\end{tabular}
\label{tab:total_complexity}
\end{table}

The key innovation replaces the quadratic $O(nm)$ search complexity with $O(n\log m)$ indexed retrieval. For large-scale datasets where $m \gg kL$ and $m \gg \log m$, this yields dramatic performance improvements. The corresponding search reduces the verification workload from $O(nmL)$ to $O(nkL)$, which enables efficient processing while maintaining reliable matching.

Asymptotic analysis confirms our superiority since:
\begin{equation}
\lim_{m \to \infty} \frac{\log m + kL}{m(1 + L)} = 0
\end{equation}
guaranteeing better scaling as dataset sizes increase. Our approach preserves full verification rigor while achieving orders-of-magnitude acceleration over brute-force methods.

\section{Conclusion}

We introduced "Transformer-Gather, Fuzzy-Reconsider," a hybrid framework that achieves scalable entity resolution through three key innovations. First, our natural language serialization enables effective semantic encoding of structured records. Second, KNN-indexed retrieval reduces the computational complexity from $O(n^2)$ to $O(n\log n)$. Third, deterministic fuzzy verification ensures high reliability while maintaining explainability.

Experimental results demonstrate that our framework outperforms brute-force methods in processing speed without compromising matching quality. The ablation studies, which were illustrated earlier, validate that both the transformer-based semantic understanding and fuzzy syntactic verification are essential components for achieving optimal performance.

Future work will focus on two primary enhancements. First, the deterministic fuzzy analysis could be replaced by a learnable string similarity model, such as a lightweight Siamese network, to capture domain-specific character variations more effectively. Second, the framework could be extended to identify complex, multi-field contradictions, moving beyond the current single-field analysis. Exploring these learned syntactic models would provide a more robust and adaptive matching capability.


\ifCLASSOPTIONcompsoc
  \section*{Acknowledgments}
\else
  \section*{Acknowledgment}
\fi

The authors gratefully acknowledge Greenweb Company for providing access to relevant databases from its subsidiary, Iranserver, as well as for supplying the scientific and technical infrastructure that made this study possible.

\vfill



\bibliographystyle{IEEEtran}
\bibliography{IEEEabrv,IEEEexample}

%




\end{document}